\documentclass[lettersize,journal]{IEEEtran}

\usepackage{hyperref}
\usepackage{amsmath,amsfonts}
\usepackage{physics,amsmath}
\usepackage{amssymb}
\usepackage{commath}
\usepackage{algorithm}
\usepackage{algpseudocode}
\usepackage{array}
\usepackage{amsmath,amssymb}
\usepackage{subcaption}
\usepackage{bm}
\usepackage{subcaption}

\usepackage{mathtools}
\usepackage{xcolor}
\usepackage{tikz}
\usetikzlibrary{arrows.meta, positioning, shapes.geometric, shadows}
\usepackage{subcaption}
\usepackage{textcomp}
\usepackage{stfloats}
\usepackage{lipsum}  
\usepackage{url}
\usepackage{verbatim}
\usepackage{graphicx}
\usepackage{cite}
\usepackage{pdfpages}
\usepackage{algorithm}
\usepackage{algpseudocode} 
\usepackage{tikz}
\usetikzlibrary{calc}
\usetikzlibrary{3d}
\usetikzlibrary{arrows.meta} 
\usepackage{amsmath,amssymb}
\usepackage{amsmath,bm}
\usepackage{tikz}
\usetikzlibrary{arrows.meta,calc,decorations.pathreplacing}
\DeclareMathOperator{\Var}{Var}
\DeclareMathOperator{\Cov}{Cov}

\DeclareCaptionLabelFormat{simple}{#2}
\newtheorem{proposition}{Proposition}
\newtheorem{theorem}{Theorem}

\begin{document}

\title{
Dispersion-Domain Detection for Mobile Molecular Communication Under Multiplicative Geometry Uncertainty}
\author{Shaojie Zhang,~\IEEEmembership{Student             Member,~IEEE}
        and Ozgur B. Akan,~\IEEEmembership{Fellow,~IEEE}
        \thanks{Shaojie Zhang is with the Internet of Everything Group, Electrical
        Engineering Division, Department of Engineering, University of Cambridge,
        CB3 0FA Cambridge, U.K. (e-mail: sz466@cam.ac.uk). }
        \thanks{Ozgur B. Akan is with the Internet of Everything Group, Electrical
        Engineering Division, Department of Engineering, University of Cambridge,
        CB3 0FA Cambridge, U.K., and also with the Center for neXt-Generation
        Communications (CXC), Department of Electrical and Electronics Engineering, Ko\text{\c{c}} University, 34450 Istanbul, Turkey (e-mail: oba21@cam.ac.uk and
        akan@ku.edu.tr)}
}


\maketitle

\begin{abstract}
Mobile molecular communication (MC) links with counting receivers are sensitive to transmitter--receiver geometry especially when nodes are mobile. We study binary detection from within-symbol count observations with unknown finite-memory inter-symbol interference (ISI) and a block-constant multiplicative geometry gain. Under a mixed-Poisson view mobility and geometry uncertainty can randomize the latent received intensity and create extra-Poisson dispersion. We propose a profiled dispersion-domain statistic $T_k^{(\Delta)}$ formed after profiling the deterministic mean shape. The statistic subtracts the intrinsic Poisson component and normalizes by the squared profiled mean to target threshold stability under the stated multiplicative-gain model. Activity gating makes conditional and gate-integrated false-alarm probabilities explicit. We characterize $T_k^{(\Delta)}$ using a time-series central-limit-theorem (CLT)-motivated Gaussian working approximation with long-run-variance dependence correction yielding Gaussian-approximate receiver operating characteristic (ROC)/bit-error-rate (BER) formulas and separability design metrics. Simulations with symbol-dependent active-Brownian mobility and finite-memory ISI support the proposed mechanism show empirical threshold stability over the tested gain range and indicate usefulness when mean-domain differences are weak unreliable or intentionally suppressed.
\end{abstract}

\section{introduction}
\noindent

Molecular communication (MC) conveys information through the emission transport and reception of chemical signals and has been studied as a candidate physical-layer technology for nanonetworks and the Internet of Bio-Nano Things (IoBNT) \cite{akyildiz2008nanonetworks,akyildiz2015iobnt,farsad2016survey}.
Diffusion-based MC is particularly relevant for molecule-counting receivers because Brownian transport gives tractable descriptions of release propagation and stochastic molecule observations \cite{pierobon2010physical,nakano2013molecular}.
The received counts are shaped by random propagation residual molecules from previous emissions external interference and the long-tailed diffusion response which introduces inter-symbol interference (ISI) and makes detection sensitive to physical operating conditions \cite{moore2009noise,noel2014unifying,pierobon2013capacity}.


For molecule-counting receivers channel state information (CSI) depends on diffusion parameters flow reactions transmitter--receiver separation receiver properties the channel impulse response (CIR) and interference statistics. Coherent detection/equalization and classical sampling-based designs therefore often rely on known or estimated CSI and can become more complex as ISI grows \cite{meng2014receiver,kilinc2013receiver,mahfuz2014sampling,noel2014optimalflow}.
When CSI is learned from data training overhead and modeling mismatch affect detection reliability \cite{jamali2016channelestimation}. Non-coherent alternatives reduce explicit CSI dependence by using non-coherent maximum-likelihood (ML) transient-feature high-dimensional or coding-based designs \cite{jamali2018noncoherent,li2020csiindependent,wei2020highdimensional,jamali2018constantcomposition}.
Mobility and uncertain geometry add a different difficulty. Transmitter--receiver separation changes can make the CIR time-varying and cause detectors calibrated for a fixed distance to fail \cite{ahmadzadeh2017mobile,chang2018adaptive}.
Related chemo-hydrodynamic transceiver models also highlight that propulsion and molecular transmission can be coupled design dimensions in IoBNT settings \cite{zhang2026chemoHydrodynamic}.
Existing mobile or uncertain-geometry responses therefore commonly estimate distance/CIR, adapt thresholds, mitigate ISI, stabilize the mean level through state-space tracking or power control, or use environment-robust signaling ideas such as equilibrium signaling \cite{chang2018adaptive,jing2022ekf,akdeniz2021equilibrium}.


In this work the receiver uses within-symbol count samples and models them as conditionally Poisson given an underlying sampled intensity process.
The key modeling point is that this latent intensity can itself be random. Bursty external release processes and unmodeled interference can randomize the rate \cite{noel2014unifying,etemadi2019cpns} and mobility or unknown separation can introduce random effective gain and time variation.
Under this mixed-Poisson/Cox view geometry and mobility fluctuations create extra-Poisson dispersion in the counts so second-order fluctuations carry information that is not captured by mean-level adaptation alone.


We therefore formulate binary mobile-MC detection as a composite-hypothesis problem with multiplicative geometry uncertainty represented by a nuisance gain.
To the best of our knowledge, the most closely related mobile or uncertain-geometry MC detectors in \cite{chang2018adaptive,jing2022ekf,akdeniz2021equilibrium} primarily operate through mean-level adaptation, CSI recovery, or equilibrium signaling, whereas the receiver developed here exploits overdispersion induced by latent intensity fluctuations.
The proposed statistic profiles the within-symbol mean shape subtracts the intrinsic Poisson contribution and normalizes the excess-dispersion term by the squared profiled mean.
This yields a dispersion-domain decision statistic designed to support threshold stability under the stated multiplicative-gain model especially when the mean level is unreliable or the channel varies across symbols due to unknown or time-varying transmitter--receiver separation.


The main contributions of this work are summarized as follows. 
\begin{itemize}
  \item We identify mobility-induced overdispersion as a usable detection signature in mobile molecular communication and cast binary count detection under unknown geometry as a composite mixed-Poisson problem with a nuisance multiplicative gain.
  \item We give the detector implementation. Mean-profile profiling a normalized excess-dispersion statistic and gate-integrated threshold calibration.
  \item We provide a dependence-aware Gaussian working analysis including receiver operating characteristic (ROC)/bit-error-rate (BER) approximations and separability metrics and validate the receiver under symbol-dependent mobility ISI and geometry uncertainty.
\end{itemize}

\section{System Model}
\label{system model}
We first define the count observation model and the nuisance-geometry structure used by both the detector and the analysis.

\begin{figure}[t]
  \centering
  \includegraphics[width=\columnwidth]{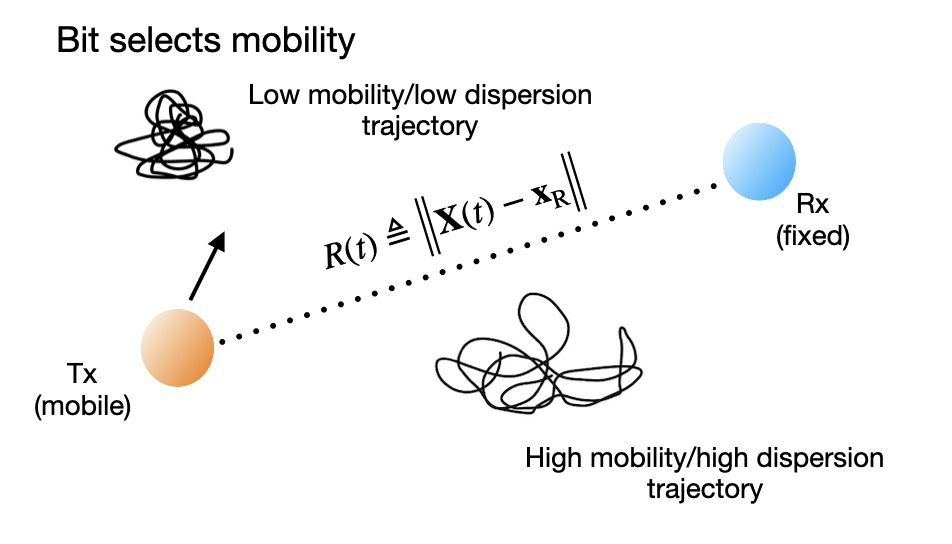}
  \caption{Conceptual overview: mobility modulation induces a random transmitter--receiver separation $R(t)$.}
  \label{fig:concept_overview}
\end{figure}

\subsection{Detector-Facing Observation Law}\label{sec:sys_physical_sampling}

Binary symbols $s_k\in\{0,1\}$ are conveyed by release amplitudes
$A(s_k)\in\{A_0,A_1\}$. Over a symbol interval of duration $T_{\mathrm{sym}}$, the receiver records the count vector
$\mathbf{Y}_k\triangleq [Y_{k,1},\ldots,Y_{k,M}]^\top$ at within-symbol offsets
$t_m=(m-\tfrac{1}{2})T_{\mathrm{sym}}/M$, with symbol start $t_k=(k-1)T_{\mathrm{sym}}$. The detector only uses this
vector, or a windowed subset of its entries.

We model the samples through the latent intensity law
\begin{align}
Y_{k,m}\mid \Lambda_{k,m}
&\sim \mathrm{Poisson}(\Lambda_{k,m}), \label{eq:cond_poiss_IIA}\\
\Lambda_{k,m}
&= \lambda_{\mathrm{bg}}+\Psi\,\widetilde{\Lambda}_{k,m},
\qquad m=1,\ldots,M . \label{eq:geom_gain_model_IIB}
\end{align}
Here $\lambda_{\mathrm{bg}}\ge 0$ is a constant background rate, $\Psi>0$ is the packet-level multiplicative geometry
gain, and
$\widetilde{\Lambda}_{k,m}$ is the gain-normalized contribution of the current release together with finite-memory ISI
from $(s_k,s_{k-1},\ldots,s_{k-L+1})$, where $L$ is the ISI memory length.

A concrete physical specialization of the gain-normalized signal component is the finite-memory superposition
\begin{equation}
\widetilde{\Lambda}_{k,m}
=
\sum_{\ell=0}^{L-1}
A(s_{k-\ell})\,h\!\left(r_{k,m},\,\ell T_{\mathrm{sym}}+t_m\right),
\label{eq:sim_tap_superposition_main}
\end{equation}
where $r_{k,m}$ is the transmitter--receiver separation at the $m$th within-symbol sampling offset and
\begin{equation}
h(r,t)=g_0(4\pi D_m t)^{-3/2}\exp\!\left(-\frac{r^2}{4D_m t}\right)u_{\mathrm{step}}(t)
\label{eq:sim_kernel_main}
\end{equation}
is the free-space passive-receiver diffusion response \cite{crank1975diffusion,pierobon2010physical,noel2014optimal}. Here
$g_0$ is an effective receiver-volume gain, $D_m$ is the molecular diffusion coefficient, and $u_{\mathrm{step}}(t)$ is the unit-step
function.

\subsection{Nuisance Gain and Mobility-Induced Dispersion}\label{sec:sys_geom_nuisance}
\label{sec:sys_mobility}

The gain $\Psi$ collects the dominant multiplicative geometry and receiver-efficiency effect. We model it as
block-constant over a packet of $K$ symbol intervals, where $K$ denotes the packet length in symbols: the same $\Psi$ applies to all samples in that packet, while
$\Psi$ may change between packets. Since $\Psi$ is unknown, the receiver targets dispersion statistics and thresholds
that remain stable over the stated multiplicative-gain range.

To motivate the random-intensity component exploited by the detector, let $X(t)\in\mathbb{R}^3$ be the transmitter
position with initial value $x_0$, $x_R$ the receiver location, $R(t)\triangleq \|X(t)-x_R\|$, and
$r_{k,m}=R(t_k+t_m)$. During symbol interval $k$, the symbol selects an active Brownian particle (ABP) mobility state
\begin{equation}\label{eq:abp_sde_IIC_3d}
\begin{aligned}
\mathrm{d}X(t)
&=v(s_k)\,n(t)\,\mathrm{d}t+\sqrt{2D_t}\,\mathrm{d}W(t),\\
&\qquad t\in[t_k,t_k+T_{\mathrm{sym}}),
\end{aligned}
\end{equation}
where $v(s_k)$ is the self-propulsion speed, $n(t)$ is a unit-norm orientation vector with rotational diffusion
coefficient $D_r(s_k)$, $D_t$ is the translational diffusion coefficient, and $W(t)$ is a standard 3-D Brownian motion
\cite{bechinger2016active,tenhagen2011brownian}. The corresponding long-time
effective diffusivity is
\begin{equation}\label{eq:Deff_3d_IIC}
D_{\mathrm{eff}}(s)=D_t+\frac{v(s)^2}{6D_r(s)},\qquad s\in\{0,1\}.
\end{equation}
Thus different choices of $(v(s),D_r(s))$ induce different separation fluctuations, which make
$\widetilde{\Lambda}_{k,m}$ random even after conditioning on the symbol sequence. Through
\eqref{eq:cond_poiss_IIA}, this randomness appears as symbol-dependent extra-Poisson dispersion in the observed counts.

\section{Observation Statistics and Dispersion}\label{sec:obs_stats}

\subsection{Mixed-Poisson / Overdispersion Identity}\label{sec:mixedpoiss_identity}

Under the conditional Poisson model in \eqref{eq:cond_poiss_IIA}, the latent intensity $\Lambda_{k,m}$ may itself be
random due to mobility- and geometry-induced fluctuations, leading to a mixed-Poisson
(Cox) observation. In this case, the unconditional variance of $Y_{k,m}$ admits the decomposition
\begin{align}
\mathrm{Var}(Y_{k,m})
&= \mathbb{E}\!\left[\mathrm{Var}\!\left(Y_{k,m}\mid \Lambda_{k,m}\right)\right]
   + \mathrm{Var}\!\left(\mathbb{E}\!\left[Y_{k,m}\mid \Lambda_{k,m}\right]\right) \nonumber\\
&= \mathbb{E}[\Lambda_{k,m}] + \mathrm{Var}(\Lambda_{k,m}),
\label{eq:mixedpoiss_overdisp_identity}
\end{align}
since $\mathbb{E}[Y_{k,m}\mid \Lambda_{k,m}]=\Lambda_{k,m}$ and $\mathrm{Var}(Y_{k,m}\mid \Lambda_{k,m})=\Lambda_{k,m}$.

Equation~\eqref{eq:mixedpoiss_overdisp_identity} separates intrinsic counting noise from dispersion induced by rate
randomness. In particular, the deviation from the Poisson law is quantified by
$\mathrm{Var}(Y_{k,m})-\mathbb{E}[Y_{k,m}]=\mathrm{Var}(\Lambda_{k,m})$, which isolates the excess variance created by
mobility/geometry uncertainty. This excess term is the statistical footprint that dispersion-domain receivers can exploit
when mean-based detection is unreliable or intentionally suppressed.

\subsection{Mean-Profile Removal and Residuals}\label{sec:mean_profile_residuals}

The raw within-symbol counts also contain a deterministic mean shape from diffusion and truncated ISI, so we remove a
per-symbol scaled template before forming the dispersion statistic. We model this mean profile as
\begin{equation}\label{eq:mu_hat_profile}
\widehat{\mu}_k(m)\triangleq \widehat{a}_k\,u(m)+\widehat{b}_k,\qquad m=1,\ldots,M,
\end{equation}
where $u(m)\ge 0$ is a fixed within-symbol template (common across $k$).
Let $\mathcal{J}\subseteq\{1,\ldots,M\}$ denote the retained set of within-symbol sample indices; $M$ remains the total
number of samples per symbol.
The per-symbol parameters $(\widehat{a}_k,\widehat{b}_k)$ are fitted on the retained indices $m\in\mathcal{J}$ by
\begin{equation}\label{eq:ab_mle}
\begin{aligned}
(\widehat{a}_k,\widehat{b}_k)
&\in\arg\min_{a\ge 0,\,b\ge 0}
\sum_{m\in\mathcal{J}}
\Big(a\,u(m)+b\\
&\qquad\qquad{}- Y_{k,m}\log\!\big(a\,u(m)+b\big)\Big).
\end{aligned}
\end{equation}
The residual used below is
\begin{equation}\label{eq:residual_def}
\widetilde{Y}_{k,m}\triangleq Y_{k,m}-\widehat{\mu}_k(m),\qquad m=1,\ldots,M,
\end{equation}
with only $m\in\mathcal{J}$ entering $T_k^{(\Delta)}$.

The template shape is learned once from calibration symbols $\mathcal{K}_{\rm cal}$:
\begin{equation}\label{eq:mu_cal}
\widehat{\mu}^{\rm cal}(m)\triangleq
\frac{1}{|\mathcal{K}_{\rm cal}|}\sum_{k\in\mathcal{K}_{\rm cal}} Y_{k,m},
\qquad m=1,\ldots,M,
\end{equation}
\begin{equation}\label{eq:u_from_cal}
u(m)\leftarrow
\frac{\widehat{\mu}^{\rm cal}(m)}
{\frac{1}{M}\sum_{j=1}^{M}\widehat{\mu}^{\rm cal}(j)} .
\end{equation}
Thus $u(m)$ captures the normalized within-symbol mean shape, while $(\widehat{a}_k,\widehat{b}_k)$ are fitted online.

After detrending, we use the standard within-slot approximation that, under each hypothesis, the windowed residual sequence
$\{\widetilde{Y}_{k,m}\}_{m\in\mathcal{J}}$ is approximately wide-sense stationary (WSS) and mixing in $m$:
\begin{equation}\label{eq:wss_on_residuals}
\mathrm{Cov}\!\big(\widetilde{Y}_{k,m},\widetilde{Y}_{k,m+\ell}\mid \mathcal{H}_s\big)\approx
\Gamma_{\widetilde{Y},s}(\ell), \qquad \ell\ge 0.
\end{equation}

\subsection{The Proposed Dispersion Statistic $T_k^{(\Delta)}$}\label{sec:Tdelta_def}

Following Sec.~III-B, we form a scale-free within-symbol dispersion statistic from the detrended samples.
For the retained index set $\mathcal{J}$, define
\begin{equation}\label{eq:Meff_def}
M_{\mathrm{eff}}\triangleq |\mathcal{J}|.
\end{equation}
Let $p$ denote the number of mean-profile parameters estimated per symbol in $\widehat{\mu}_k(m)$ (e.g., $p=2$ for a
scale-and-offset fit). For each $m\in\mathcal{J}$, define the normalized excess-dispersion contribution
\begin{equation}\label{eq:psi_hat_def}
\widehat{\psi}_{k,m}\triangleq
\frac{\big(Y_{k,m}-\widehat{\mu}_k(m)\big)^2 - Y_{k,m}}
{\widehat{\mu}_k(m)^{2}},
\end{equation}
and compute
\begin{equation}\label{eq:Tdelta_def_final}
T_k^{(\Delta)}\triangleq
\frac{1}{M_{\mathrm{eff}}-p}\sum_{m\in\mathcal{J}}\widehat{\psi}_{k,m}.
\end{equation}

The numerator in \eqref{eq:psi_hat_def} removes the intrinsic Poisson component, so $T_k^{(\Delta)}$ aggregates
within-symbol dispersion beyond pure counting noise, consistent with the mixed-Poisson identity in
Sec.~\ref{sec:mixedpoiss_identity}.
The superscript $(\Delta)$ is a statistic label for this excess-dispersion construction, not an additional tuning
parameter.
The normalization by $\widehat{\mu}_k(m)^2$ implements the scale adjustment used by the detector under the stated
geometry/mobility model. In the simulations, $\mathcal{J}$ is chosen by a template-energy rule. The calibration details
are given in Appendix~\ref{app:window_and_dependence}.

\section{Detector Design}\label{sec:detector_design}

\subsection{Hypotheses and Composite Nature}\label{sec:det_hypotheses}

Using the within-symbol samples $\mathbf{Y}_k=[Y_{k,1},\ldots,Y_{k,M}]^\top$, the receiver performs binary detection of the $k$th symbol
\begin{equation}\label{eq:IV_H0H1}
\mathcal{H}_0:\ s_k=0
\qquad \text{vs.}\qquad
\mathcal{H}_1:\ s_k=1 .
\end{equation}
Throughout, the decision is based on $\mathbf{Y}_k$ (or a scalar statistic computed from it), while the underlying
sampling law follows the mixed-Poisson model introduced in Sec.~II.

A key complication is that the distribution of $\mathbf{Y}_k$ under either hypothesis is not specified uniquely by $s_k$.
Instead, it depends on additional quantities that are not assumed known at the receiver. In particular, the mean level
and within-slot mean shape inherit (i) an unknown multiplicative geometry gain $\Psi$ and background level
$\lambda_{\mathrm{bg}}$ as defined in \eqref{eq:geom_gain_model_IIB}, and (ii) an unknown interference state from the preceding
symbols through the finite-memory ISI superposition (length $L$) in the channel model. For compactness, we collect these
unknowns into a nuisance vector
\begin{equation}\label{eq:IV_nuisance_vec}
\boldsymbol{\nu}_k \triangleq \big(\Psi,\lambda_{\mathrm{bg}},\mathbf{s}_{k-1}\big),
\qquad \mathbf{s}_{k-1}\triangleq (s_{k-1},\ldots,s_{k-L+1}),
\end{equation}
noting that $\boldsymbol{\nu}_k$ is present under both $\mathcal{H}_0$ and $\mathcal{H}_1$.
Therefore, symbol detection is naturally a composite hypothesis testing problem: the receiver must decide between
$\mathcal{H}_0$ and $\mathcal{H}_1$ in the presence of unknown nuisance parameters that affect the likelihood under each
hypothesis.

This composite viewpoint motivates the scale-normalized detector and the nuisance-profiled baselines below.

\subsection{Threshold-Stable Dispersion Detector}\label{sec:det_threshold_stability}
We consider a dispersion-domain detector based on the scale-normalized statistic
$T_k^{(\Delta)}$ in \eqref{eq:Tdelta_def_final}. The per-symbol decision rule is a single-threshold test
\begin{equation}\label{eq:Tk_threshold_rule}
T_k^{(\Delta)}
\mathop{\gtrless}_{\mathcal{H}_0}^{\mathcal{H}_1}
\tau_T ,
\end{equation}
where $\tau_T$ is calibrated once and then reused across packets with different multiplicative geometry gains.
The normalization embedded in $T_k^{(\Delta)}$ through \eqref{eq:psi_hat_def}--\eqref{eq:Tdelta_def_final}
targets threshold stability under multiplicative geometry uncertainty. The population cancellation is exact only under
pure multiplicative scaling of the signal component. With profiled means, additive background, gating, and finite samples,
the implemented detector should be interpreted through threshold stability over the stated gain regime.

We set $\tau_T$ using labeled $\mathcal{H}_0$ symbols only.
Let $\mathcal{K}_0$ denote a set of calibration symbols known to satisfy $\mathcal{H}_0$.
If a light activity gate is used to avoid ill-conditioned mean-profile fitting at extremely low count levels, define
\begin{equation}\label{eq:gate_def}
\mathcal{G}_k \triangleq \big\{\bar{Y}_{k,\mathcal{J}}>\tau_Y\big\},\qquad
\bar{Y}_{k,\mathcal{J}}\triangleq \frac{1}{M_{\mathrm{eff}}}\sum_{m\in\mathcal{J}} Y_{k,m},
\end{equation}
where $\tau_Y\ge 0$ is a fixed gate threshold.
A convenient gate choice is to calibrate $\tau_Y$ as an $\alpha_{\mathrm{gate}}$-quantile of $\bar{Y}_{k,\mathcal{J}}$
under $\mathcal{H}_0$, so that $ \Pr(\mathcal{G}_k\mid \mathcal{H}_0)\approx 1-\alpha_{\mathrm{gate}}$ in calibration.

For a target conditional false-alarm level $P_{\mathrm{FA}}^\star \triangleq \Pr(T_k^{(\Delta)}>\tau_T\mid\mathcal{H}_0,\mathcal{G}_k)$,
we choose $\tau_T$ as the empirical $(1-P_{\mathrm{FA}}^\star)$-quantile of the gated calibration realizations
$\{T_k^{(\Delta)}: k\in\mathcal{K}_0,\ \mathcal{G}_k\}$:
\begin{equation}\label{eq:tauT_quantile}
\tau_T \leftarrow
\widehat{F}^{-1}_{T\mid \mathcal{H}_0,\mathcal{G}}
\big(1-P_{\mathrm{FA}}^\star\big),
\end{equation}
where $\widehat{F}_{T\mid \mathcal{H}_0,\mathcal{G}}$ is the empirical CDF of $T_k^{(\Delta)}$ under $\mathcal{H}_0$
conditioned on $\mathcal{G}_k$.
If the receiver never declares $\widehat{s}_k=1$ when the gate fails, i.e., $\bar{Y}_{k,\mathcal{J}}\le\tau_Y$ implies
$\widehat{s}_k\neq 1$, then the gate-integrated, unconditional false-alarm probability is
\begin{equation}\label{eq:Pfa_uncond_gate}
P_{\mathrm{FA}}
\triangleq \Pr(\widehat{s}_k=1\mid\mathcal{H}_0)
=
\Pr(\mathcal{G}_k\mid\mathcal{H}_0)\,P_{\mathrm{FA}}^\star .
\end{equation}
Thus the reported false-alarm probability includes both the activity gate and the conditional dispersion-threshold event.

\begin{algorithm}[t]
\caption{Threshold-stable dispersion detector (per symbol $k$)}
\label{alg:threshold_stable_detector}
\begin{algorithmic}[1]
\Require Window $\mathcal{J}$, template $u(m)$, thresholds $(\tau_Y,\tau_T)$, and statistic definitions
\eqref{eq:mu_hat_profile}--\eqref{eq:Tdelta_def_final}.
\State (Gate) Compute $\bar{Y}_{k,\mathcal{J}}$. If $\bar{Y}_{k,\mathcal{J}}\le \tau_Y$, declare $\widehat{s}_k=0$ and stop.
\State (Template removal) Fit $(\widehat{a}_k,\widehat{b}_k)$ on $m\in\mathcal{J}$ and form $\widehat{\mu}_k(m)$ via \eqref{eq:mu_hat_profile}.
\State (Dispersion statistic) Compute $\widehat{\psi}_{k,m}$ and $T_k^{(\Delta)}$ via \eqref{eq:psi_hat_def}--\eqref{eq:Tdelta_def_final}.
\State (Decision) Output $\widehat{s}_k=1$ if $T_k^{(\Delta)}>\tau_T$, else $\widehat{s}_k=0$.
\end{algorithmic}
\end{algorithm}

\subsection{Baselines and Benchmarks}\label{sec:det_baselines}

To contextualize the empirical behavior of the dispersion-domain rule in
Sec.~\ref{sec:det_threshold_stability}, we report three representative comparison
receivers. All methods operate on the same within-symbol sample set and
window $\mathcal{J}$ (hence the same $M_{\mathrm{eff}}$), and apply the same
gate $\mathcal{G}_k$ in \eqref{eq:gate_def} so that gate failures are treated
uniformly as ``no-alarm'' decisions. Throughout, we use equiprobable symbols
so the likelihood-ratio threshold is $\eta=1$.

\medskip
\noindent\textbf{1) Mean/energy baseline.}
Using the windowed mean $\bar{Y}_{k,\mathcal{J}}$ in \eqref{eq:gate_def}, the
mean-based detector is the single-threshold test
\begin{equation}\label{eq:baseline_mean_rule}
\bar{Y}_{k,\mathcal{J}}
\mathop{\gtrless}_{\mathcal{H}_0}^{\mathcal{H}_1}
\tau_{\bar{Y}},
\end{equation}
where $\tau_{\bar{Y}}$ is calibrated under $\mathcal{H}_0$ at the same target
false-alarm level used for $T_k^{(\Delta)}$.

\medskip
\noindent\textbf{2) Genie-aided oracle benchmark (known nuisance).}
As an upper benchmark, assume the nuisance vector
$\boldsymbol{\nu}_k=(\Psi,\lambda_{\mathrm{bg}},\mathbf{s}_{k-1})$ is known.
Let $\mathbf{Y}_{k,\mathcal{J}}\triangleq \{Y_{k,m}\}_{m\in\mathcal{J}}$.
The oracle likelihood-ratio test is
\begin{equation}\label{eq:baseline_oracle_LRT}
\Lambda_k^{\mathrm{orc}}(\mathbf{Y}_{k,\mathcal{J}})
\triangleq
\frac{p(\mathbf{Y}_{k,\mathcal{J}}\mid \mathcal{H}_1,\boldsymbol{\nu}_k)}
     {p(\mathbf{Y}_{k,\mathcal{J}}\mid \mathcal{H}_0,\boldsymbol{\nu}_k)}
\mathop{\gtrless}_{\widehat{s}_k=0}^{\widehat{s}_k=1}
\eta,
\end{equation}
where $p(\cdot\mid\mathcal{H}_s,\boldsymbol{\nu}_k)$ is the marginal mixed-Poisson
likelihood induced by \eqref{eq:cond_poiss_IIA} and the intensity model
\eqref{eq:geom_gain_model_IIB} (i.e., it integrates out the latent intensity
randomness under the symbol-dependent mobility).

\medskip
\noindent\textbf{3) Generalized likelihood ratio test (GLRT that profiles geometry gain/background).}
To remove the unknown scale parameters while retaining a likelihood-based
comparison, we also consider the generalized likelihood ratio test that profiles
$(\Psi,\lambda_{\mathrm{bg}})$ under each hypothesis:
\begin{equation}\label{eq:baseline_GLRT}
\begin{aligned}
\Lambda_k^{\mathrm{glrt}}(\mathbf{Y}_{k,\mathcal{J}})
&\triangleq
\frac{
\displaystyle \sup_{\Psi>0,\ \lambda_{\mathrm{bg}}\ge 0}
\, p(\mathbf{Y}_{k,\mathcal{J}}\mid \mathcal{H}_1,\Psi,\lambda_{\mathrm{bg}},\widehat{\mathbf{s}}_{k-1})
}{
\displaystyle \sup_{\Psi>0,\ \lambda_{\mathrm{bg}}\ge 0}
\, p(\mathbf{Y}_{k,\mathcal{J}}\mid \mathcal{H}_0,\Psi,\lambda_{\mathrm{bg}},\widehat{\mathbf{s}}_{k-1})
}
\\
&\mathop{\gtrless}_{\widehat{s}_k=0}^{\widehat{s}_k=1}\eta .
\end{aligned}
\end{equation}
where $\widehat{\mathbf{s}}_{k-1}$ denotes the ISI side information used by the
receiver (either genie-provided past symbols or tentative decisions, as described
next). The optimization in \eqref{eq:baseline_GLRT} is low-dimensional because
geometry enters the observation law only through the gain $\Psi$ and the
(background) offset $\lambda_{\mathrm{bg}}$.

When evaluating ISI stress tests, we additionally report decision-feedback equalizer (DFE) versions of the
above baselines and of the proposed $T_k^{(\Delta)}$ rule. An initial symbol-by-symbol pass produces tentative decisions, which are then used to form $\widehat{\mathbf{s}}_{k-1}$ and suppress the dominant ISI contribution before a
second detection pass. For fairness, the same ISI memory length $L$ and the same
sampling window $\mathcal{J}$ are used across all methods.

\section{Performance Characterization}
Having specified the implemented statistic and comparison receivers, we next characterize the distribution and threshold behavior of $T_k^{(\Delta)}$.

\subsection{Working Distribution of the Implemented Statistic}\label{sec:perf_Tdelta_dist}

The implemented statistic $T_k^{(\Delta)}$ in \eqref{eq:Tdelta_def_final} is an average of the profiled contributions
$\{\widehat{\psi}_{k,m}\}_{m\in\mathcal{J}}$ in \eqref{eq:psi_hat_def}. Under the within-slot residual mixing/WSS
approximation in \eqref{eq:wss_on_residuals}, a time-series central limit theorem (CLT) motivates a Gaussian
large-sample characterization for moderate/large $M_{\mathrm{eff}}$.

Define the hypothesis-conditioned dispersion level
\begin{equation}\label{eq:delta_s_def_VA}
\delta_s \triangleq \mathbb{E}\!\left[T_k^{(\Delta)}\mid \mathcal{H}_s\right],
\qquad s\in\{0,1\}.
\end{equation}
Let $\omega_{\psi,s}^2$ denote the finite long-run variance (LRV) of
$\{\widehat{\psi}_{k,m}\}_{m\in\mathcal{J}}$ under $\mathcal H_s$, including the effect of residual dependence across
within-symbol samples. We use the Gaussian working approximation
\begin{equation}\label{eq:Tk_gauss_VA}
T_k^{(\Delta)}\mid \mathcal{H}_s\ \dot{\sim}\ 
\mathcal{N}\!\Big(\mu_T(\mathcal{H}_s),\,\sigma_T^2(\mathcal{H}_s)\Big),
\qquad s\in\{0,1\},
\end{equation}
where $\dot{\sim}$ denotes an approximation rather than an exact finite-sample law, with
\begin{equation}\label{eq:mu_sigma_T_VA}
\begin{aligned}
\mu_T(\mathcal{H}_s)&= \delta_s,\\
\sigma_T^2(\mathcal{H}_s)&\approx \frac{M_{\mathrm{eff}}}{(M_{\mathrm{eff}}-p)^2}\,\omega_{\psi,s}^2,
\qquad s\in\{0,1\}.
\end{aligned}
\end{equation}
When the profiled contribution sequence is approximately uncorrelated across $m$,
$\omega_{\psi,s}^2=\Var(\widehat{\psi}_{k,m}\mid\mathcal{H}_s)$ and the variance in
\eqref{eq:mu_sigma_T_VA} reduces to the usual $1/M_{\mathrm{eff}}$ scaling, up to the fixed
degrees-of-freedom correction. The quantities $(\delta_s,\omega_{\psi,s}^2)$ are estimated from labeled data for the
analytical overlays. The LRV estimator is specified in Appendix~\ref{app:window_and_dependence}.

\subsection{Threshold Stability Under Multiplicative Gain}\label{sec:perf_gain_scaling}
\label{sec:perf_oracle_gain_stability}

The scale normalization in $T_k^{(\Delta)}$ is anchored by an exact population cancellation under pure multiplicative
signal scaling. After background has been removed or absorbed, suppose
$\Lambda_{k,m}=\Psi\,\widetilde{\Lambda}_{k,m}$, where the law of $\widetilde{\Lambda}_{k,m}$ is independent of $\Psi$
for a fixed hypothesis. Let
$\widetilde{\mu}_s(m)\triangleq \mathbb{E}[\widetilde{\Lambda}_{k,m}\mid\mathcal{H}_s]$ and
$\widetilde{v}_s(m)\triangleq \Var(\widetilde{\Lambda}_{k,m}\mid\mathcal{H}_s)$.

Under the mixed-Poisson model $Y_{k,m}\,|\,\Lambda_{k,m}\sim\mathrm{Poisson}(\Lambda_{k,m})$,
\begin{equation}\label{eq:VB_moments}
\begin{aligned}
\mu_{Y,s}(m)
&\triangleq \mathbb{E}\!\left[Y_{k,m}\mid \mathcal{H}_s\right]  \\
&= \mathbb{E}\!\left[\Lambda_{k,m}\mid \mathcal{H}_s\right]
 = \Psi\,\widetilde{\mu}_s(m),\\[0.25em]
\sigma^2_{Y,s}(m)
&\triangleq \Var\!\left(Y_{k,m}\mid \mathcal{H}_s\right) \\
&= \mathbb{E}\!\left[\Lambda_{k,m}\mid \mathcal{H}_s\right]
 + \Var\!\left(\Lambda_{k,m}\mid \mathcal{H}_s\right) \\
&= \Psi\,\widetilde{\mu}_s(m)+\Psi^2\,\widetilde{v}_s(m),\\[0.25em]
\sigma^2_{Y,s}(m)-\mu_{Y,s}(m)
&= \Psi^2\,\widetilde{v}_s(m).
\end{aligned}
\end{equation}
Therefore the normalized excess-dispersion level
\begin{equation}\label{eq:VB_phi_gain_independent}
\phi_s(m)\triangleq
\frac{\Var(Y_{k,m}\mid\mathcal{H}_s)-\mathbb{E}[Y_{k,m}\mid\mathcal{H}_s]}
{\mathbb{E}[Y_{k,m}\mid\mathcal{H}_s]^2}
=
\frac{\widetilde{v}_s(m)}{\widetilde{\mu}_s(m)^2},
\end{equation}
is independent of $\Psi$. Since $T_k^{(\Delta)}$ estimates this normalized excess dispersion over
$m\in\mathcal{J}$ through \eqref{eq:psi_hat_def}--\eqref{eq:Tdelta_def_final}, the identity in
\eqref{eq:VB_phi_gain_independent} explains the threshold stability targeted by the detector under the stated
multiplicative-gain model.

For an oracle statistic with the true mean profile, the corresponding contribution has gain-independent mean and
nonzero-lag covariances, and the remaining lag-zero gain dependence is a finite-count effect. The implemented statistic
replaces the true mean by the profiled Poisson fit in \eqref{eq:ab_mle}. This profiling generally leaves a first-order
score term unless an additional small-sensitivity condition holds. Appendix~\ref{app:oracle_gain_stability} gives the
exact oracle identities and the implemented-statistic expansion. Thus the gain cancellation is exact at the population
and oracle levels under pure multiplicative scaling, while threshold stability for the implemented receiver remains an
approximate property in finite samples with profiling, background, gating, and residual dependence.

\subsection{Analytical ROC and BER Approximations}\label{sec:perf_ROC_BER}

The Gaussian characterization in \eqref{eq:Tk_gauss_VA} gives closed-form approximations for the single-threshold rule
\eqref{eq:Tk_threshold_rule}. Let
$\kappa_T\triangleq \mathrm{sign}(\mu_T(\mathcal{H}_1)-\mu_T(\mathcal{H}_0))$ and use the equivalent rule
$\kappa_T T_k^{(\Delta)}\gtrless \kappa_T\tau_T$, which does not assume which hypothesis has the larger statistic.
With $Q(x)\triangleq (2\pi)^{-1/2}\int_x^\infty e^{-t^2/2}\,dt$, the resulting approximations are
\begin{equation}\label{eq:roc_ber_gaussian_block}
\begin{aligned}
P_{\mathrm{FA}}(\tau_T)
&\approx
Q\!\left(\kappa_T\frac{\tau_T-\mu_T(\mathcal{H}_0)}{\sigma_T(\mathcal{H}_0)}\right),\\
P_{\mathrm{D}}(\tau_T)
&\approx
Q\!\left(\kappa_T\frac{\tau_T-\mu_T(\mathcal{H}_1)}{\sigma_T(\mathcal{H}_1)}\right),\\
\tau_T(P_{\mathrm{FA}}^\star)
&\approx
\mu_T(\mathcal{H}_0)+
\kappa_T\sigma_T(\mathcal{H}_0)Q^{-1}(P_{\mathrm{FA}}^\star),\\
P_b(\tau_T)
&\approx
\tfrac{1}{2}\big(P_{\mathrm{FA}}(\tau_T)+1-P_{\mathrm{D}}(\tau_T)\big),
\end{aligned}
\end{equation}
where the last line assumes equiprobable bits with
$\mathcal{H}_0\leftrightarrow s_k=0$ and $\mathcal{H}_1\leftrightarrow s_k=1$.
When the activity gate is enabled, \eqref{eq:roc_ber_gaussian_block} describes the gated dispersion-threshold event,
and the gate-integrated false-alarm probability follows \eqref{eq:Pfa_uncond_gate}.

\subsection{Gaussian Separability Design Metrics}\label{sec:perf_exponents}

For design comparisons, the same scalar Gaussian model yields compact separability metrics after compression to
$T_k^{(\Delta)}$. These metrics apply to the scalar statistic rather than to the original vector observation. Write
\begin{equation}\label{eq:VD_ms_vs_def}
T_k^{(\Delta)}\mid\mathcal{H}_s \ \dot{\sim}\ \mathcal{N}(m_s,v_s),\qquad
m_s\triangleq \mu_T(\mathcal{H}_s),\ \ v_s\triangleq \sigma_T^2(\mathcal{H}_s),
\end{equation}
and define $\Delta m\triangleq m_1-m_0$. Under this scalar Gaussian model, the minimum error probability based on
$T_k^{(\Delta)}$ with equal priors is
$P_e^\star(T)=\tfrac{1}{2}\int \min\{f_0(t),f_1(t)\}\,dt$, and satisfies the Chernoff bound
$P_e^\star(T)\le \tfrac{1}{2}\exp(-D_{\mathrm{C}})$, where $D_{\mathrm{C}}$ is the Chernoff information between
$f_0=\mathcal{N}(m_0,v_0)$ and $f_1=\mathcal{N}(m_1,v_1)$. Specifically,
\begin{equation}\label{eq:VD_Chernoff_def_compact}
\begin{aligned}
D_{\mathrm{C}} &\triangleq \max_{a\in[0,1]} D(a),\\
D(a) &\triangleq \tfrac{1}{2}\log\!\Big(\tfrac{v_a}{v_0^{a}v_1^{1-a}}\Big)
+\tfrac{a(1-a)}{2}\tfrac{\Delta m^2}{v_a},
\end{aligned}
\end{equation}
with $v_a\triangleq a v_0+(1-a)v_1$.
The Bhattacharyya distance is the closed-form case $a=\tfrac{1}{2}$,
\begin{equation}\label{eq:VD_DB_def_compact}
D_{\mathrm{B}} \triangleq D(\tfrac{1}{2})
=\tfrac{1}{2}\log\!\Big(\tfrac{v_0+v_1}{2\sqrt{v_0v_1}}\Big)
+\tfrac{\Delta m^2}{4(v_0+v_1)} ,
\end{equation}
which gives $P_e^\star(T)\le \tfrac{1}{2}\exp(-D_{\mathrm{B}})$. We also use the symmetric KL divergence
$D_{\mathrm{KL}}\triangleq \tfrac{1}{2}(\mathrm{KL}_{01}+\mathrm{KL}_{10})$, which becomes
\begin{equation}\label{eq:VD_DKLsym_def_compact}
D_{\mathrm{KL}}
=\tfrac{1}{4}\!\Big(\tfrac{v_0}{v_1}+\tfrac{v_1}{v_0}-2\Big)
+\tfrac{\Delta m^2}{4}\!\Big(\tfrac{1}{v_0}+\tfrac{1}{v_1}\Big).
\end{equation}

The metrics above show how retained-window length, mobility contrast, and ISI enter the scalar-statistic separation.
From \eqref{eq:mu_sigma_T_VA}, $v_s$ decays approximately as $1/M_{\mathrm{eff}}$, with the
$(M_{\mathrm{eff}}-p)$ correction and any LRV inflation due to residual dependence. In the common regime
$v_0\approx v_1\approx v$ and $M_{\mathrm{eff}}\gg p$, the leading term in \eqref{eq:VD_DB_def_compact} gives
\begin{equation}\label{eq:VD_scaling_Meff_compact}
D_{\mathrm{B}} \approx \tfrac{\Delta m^2}{8v}
\ \propto\ M_{\mathrm{eff}}\,\Delta m^2
\ \approx\ M_{\mathrm{eff}}\,(\delta_1-\delta_0)^2 ,
\end{equation}
so the design metric grows linearly with $M_{\mathrm{eff}}$ and quadratically with the scalar dispersion contrast in
this approximation. When the dispersion gap follows the mobility contrast through $D_{\mathrm{eff}}(s)$, this gives the
design-time scaling heuristic
$\Theta\!\big(M_{\mathrm{eff}}(D_{\mathrm{eff}}(1)-D_{\mathrm{eff}}(0))^2\big)$.

ISI enters through $(m_s,v_s)$ by reducing the per-symbol contrast and by increasing residual correlation. To summarize
tap-level ISI in one scalar, define
$h_{\ell,m}\triangleq h\!\big(r_{\mathrm{ref}},\,\ell T_{\mathrm{sym}}+t_m\big)$, where $h(r,t)$ is the sampled diffusion
response and $r_{\mathrm{ref}}$ is a reference separation such as the calibration geometry used to learn the mean
template. A convenient ISI severity index is
\begin{equation}\label{eq:VD_rhoISI_def_compact}
\rho_{\mathrm{ISI}} \triangleq \frac{\sum_{\ell=1}^{L-1}\bar h_\ell}{\bar h_0},\qquad
\bar h_\ell \triangleq \frac{1}{M_{\mathrm{eff}}}\sum_{m\in\mathcal{J}} h_{\ell,m},
\end{equation}
under which increasing $\rho_{\mathrm{ISI}}$ tends to decrease $\Delta m$ and increase $(v_0,v_1)$, thereby reducing the
Gaussian separability metrics in \eqref{eq:VD_Chernoff_def_compact}--\eqref{eq:VD_DKLsym_def_compact}. These metrics
are used below as design summaries for the compressed statistic, not as fundamental limits for the full observation
vector.

\section{Simulation Methodology}\label{sec:sim_method}

To connect the analysis to finite-sample behavior, we evaluate performance using a packet-level simulator that instantiates the random transmitter--receiver separation sequence $\{r_{k,m}\}$ and then samples counts from the conditional counting law in \eqref{eq:cond_poiss_IIA}. Specifically, for each symbol interval, we propagate the symbol-dependent mobility model in \eqref{eq:abp_sde_IIC_3d} by Euler--Maruyama with step $\Delta t_{\mathrm{traj}}$ and record separations at the within-symbol sampling offsets $\{t_m\}$. These separations are inserted into the finite-memory physical specialization in \eqref{eq:sim_tap_superposition_main}, truncated to memory length $L$. Unknown geometry is emulated by a packet-constant multiplicative scale applied to the signal component only, while the background term remains unchanged. In the simulations, one run corresponds to one packet with fixed geometry scale. A small finite-size regularization $r_{k,m}\leftarrow \max\{r_{k,m},r_{\min}\}$, with $r_{\min}$ denoting the separation floor, is used before evaluating the kernel to avoid unphysical near-zero separations. We adopt the standard conditional-Poisson sampling approximation at the offsets $\{t_m\}$, while residual within-symbol dependence is accounted for in the receiver characterization. The above micro-scale regime is consistent with common MC simulation settings, e.g.,
$r_r=5~\mu$m, $r_0=10~\mu$m, $D=79.4~\mu\mathrm{m}^2/\mathrm{s}$ in \cite{deng2016_adsorption},
and $R=10~\mu$m, $D=100~\mu\mathrm{m}^2/\mathrm{s}$ in \cite{wang2015_distance}.

\begin{table}[t]
\centering
\caption{Baseline physical/mobility parameters (unless stated otherwise).}\label{tab:sim_base}
\scriptsize
\setlength{\tabcolsep}{4pt}
\renewcommand{\arraystretch}{1.05}
\begin{tabular}{@{}l c@{}}
\hline
Parameter & Value \\
\hline
Initial separation $\|x_0-x_R\|$ & $10~\mu\mathrm{m}$ \\
Receiver radius $a_{\mathrm{rx}}$ & $5~\mu\mathrm{m}$ \\
Molecular diffusion $D_m$ & $10^{-10}~\mathrm{m}^2/\mathrm{s}$ \\
Background $\lambda_{\mathrm{bg}}$ & $2$ \\
Separation floor $r_{\min}$ & $0.8~\mu\mathrm{m}$ \\
ABP $(v(0),v(1))$ & $(0,\,30)~\mu\mathrm{m/s}$ \\
ABP $(D_r(0),D_r(1))$ & $(8,\,0.8)~\mathrm{s}^{-1}$ \\
ABP $D_t$ & $2\times10^{-13}~\mathrm{m}^2/\mathrm{s}$ \\
Trajectory step $\Delta t_{\mathrm{traj}}$ & $10^{-3}~\mathrm{s}$ \\
\hline
\end{tabular}
\end{table}

\section{Result and discussion}
The numerical results are organized around four questions: whether mobility creates a measurable dispersion signature, whether the threshold is stable under gain variation, how mobility contrast affects BER, and how residual dependence and ISI change performance.

\subsection{Dispersion Evidence}
\begin{figure*}[ht] 
     \centering
     \begin{subfigure}[b]{0.33\textwidth}
         \centering
         \includegraphics[width=\textwidth]{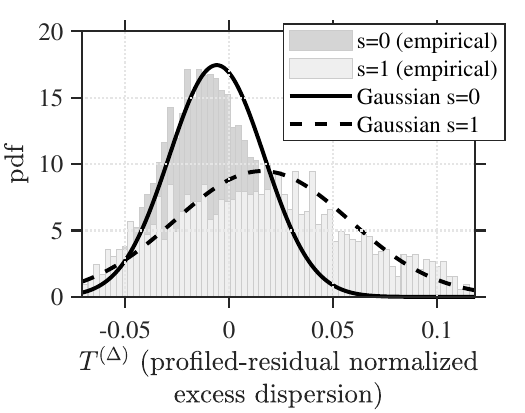}
         \caption{}
         \label{f1}
     \end{subfigure}
     \hfill 
     \begin{subfigure}[b]{0.33\textwidth}
         \centering
         \includegraphics[width=\textwidth]{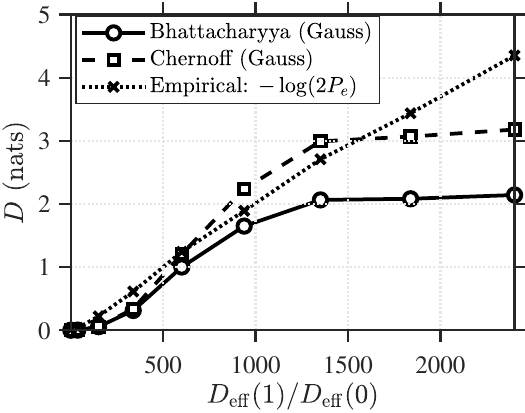}
         \caption{}
         \label{f7a}
     \end{subfigure}
     \hfill
     \begin{subfigure}[b]{0.33\textwidth}
         \centering
         \includegraphics[width=\textwidth]{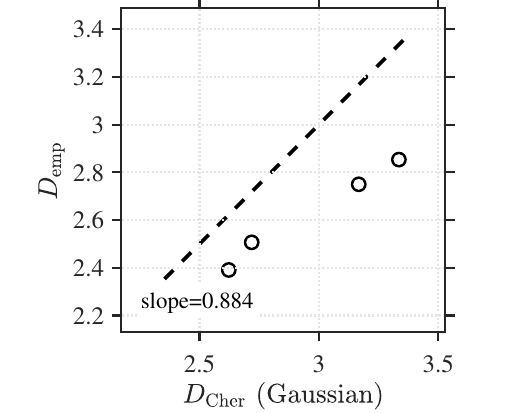}
         \caption{}
         \label{f7c}
     \end{subfigure}
     
     \caption{Dispersion-domain evidence and reliability trends: (a) empirical distributions of $T^{(\Delta)}$ with Gaussian overlays, (b) Gaussian separability metric versus mobility contrast, and (c) empirical--Gaussian comparison of the metric.}
     \label{fig:three_graphs}
\end{figure*}
\begin{figure*}[ht] 
     \centering
     \begin{subfigure}[b]{0.45\textwidth}
         \centering
         \includegraphics[width=\textwidth]{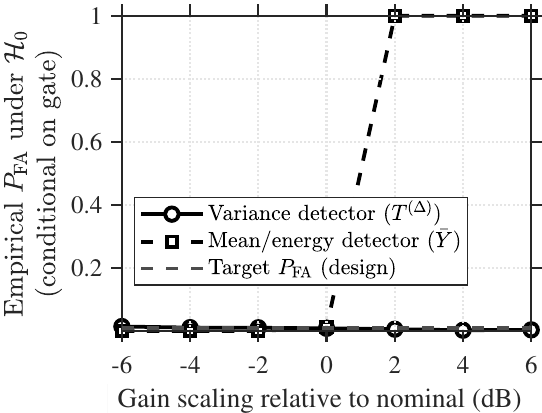}
         \caption{}
         \label{f2}
     \end{subfigure}
     \hfill 
     \begin{subfigure}[b]{0.45\textwidth}
         \centering
         \includegraphics[width=\textwidth]{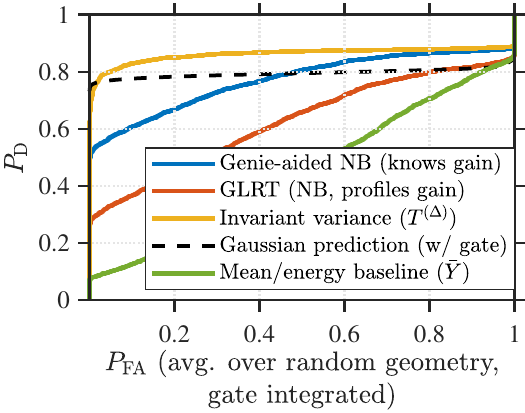}
         \caption{}
         \label{f3}
     \end{subfigure}
     \caption{Empirical threshold stability over the tested multiplicative gain range. (a) Empirical $P_{\mathrm{FA}}$ under $\mathcal{H}_0$ versus multiplicative gain scaling for $T_k^{(\Delta)}$. (b) Gate-integrated ROC averaged over random geometry-gain realizations.}
     \label{ThresholdStability}
\end{figure*}

Fig.~\ref{fig:three_graphs} provides evidence that the proposed profiled-residual dispersion statistic
$T_k^{(\Delta)}$ carries hypothesis information in the tested mixed-Poisson regime.
In Fig.~\ref{fig:three_graphs}(a), the empirical distributions under $\mathcal{H}_0$ and $\mathcal{H}_1$ are separated,
with $\mathcal{H}_1$ shifted toward larger normalized excess dispersion, consistent with $\delta_1>\delta_0$ in the
Gaussian working model for $T_k^{(\Delta)}$.
This agrees with the intended mechanism: symbol-dependent mobility randomizes the latent intensity and increases
extra-Poisson variability beyond pure counting noise. The Gaussian overlays capture the main empirical trend under
both hypotheses, consistent with their use as working approximations for moment characterization and performance prediction.

Fig.~\ref{fig:three_graphs}(b) links dispersion contrast to reliability through Gaussian separability design metrics. The Bhattacharyya/Chernoff-type separability
measures increase with mobility contrast $D_{\mathrm{eff}}(1)/D_{\mathrm{eff}}(0)$, and the empirical proxy $-\log(2\widehat{P}_b)$, with $\widehat{P}_b$ denoting the measured BER,
follows the same trend. This monotonicity is consistent with the performance scaling: for $v_0\approx v_1$ and
$M_{\mathrm{eff}}\gg p$, the dominant exponent term grows approximately linearly with $M_{\mathrm{eff}}$ and quadratically
with the dispersion gap $\Delta m\approx \delta_1-\delta_0$.

Finally, Fig.~\ref{fig:three_graphs}(c) compares empirical reliability with the Gaussian Chernoff surrogate and shows
a roughly linear trend over the tested operating points. Although finite-sample effects prevent tightness,
this is consistent with using Gaussian separability metrics as design-time predictors for
mobility-modulated signaling based on $T_k^{(\Delta)}$.

\subsection{Threshold Stability and ROC}

Fig.~\ref{ThresholdStability} evaluates threshold stability under multiplicative geometry uncertainty by comparing the proposed dispersion-domain
test with mean-based baselines and gain-aware references under matched window and gate settings. The proposed test does
not know the packet gain. The GLRT profiles the gain/background parameters. The genie-aided benchmark is supplied the
nuisance vector used in Sec.~\ref{sec:det_baselines}. In Fig.~\ref{ThresholdStability}(a), the gate-integrated empirical false-alarm
rate under $\mathcal{H}_0$ stays close to the design target over the tested multiplicative gain scalings for $T_k^{(\Delta)}$,
indicating reduced sensitivity to packet-level amplitude changes. In contrast, the mean/energy detector $\bar{Y}$
shows strong gain dependence, with false alarms rising sharply as gain increases, since it aggregates raw counts whose
level tracks the unknown scale.

Fig.~\ref{ThresholdStability}(b) shows the corresponding gate-integrated ROC averaged over random geometry-gain realizations. The proposed dispersion
detector is close to the gain-profiled GLRT over the tested regime and remains below the genie-aided benchmark, while
outperforming $\bar{Y}$ in this setting. The residual gap to the genie curve is consistent with $T_k^{(\Delta)}$ retaining primarily
second-order (dispersion) information, whereas gain-aware likelihood tests can additionally exploit any remaining
mean-shape differences after profiling. The dashed Gaussian prediction, computed with the same gate as the empirical evaluation,
captures the ROC trend of $T_k^{(\Delta)}$.

\subsection{Mobility Contrast}
\begin{figure}[t]
  \centering
  \includegraphics[width=0.9\columnwidth]{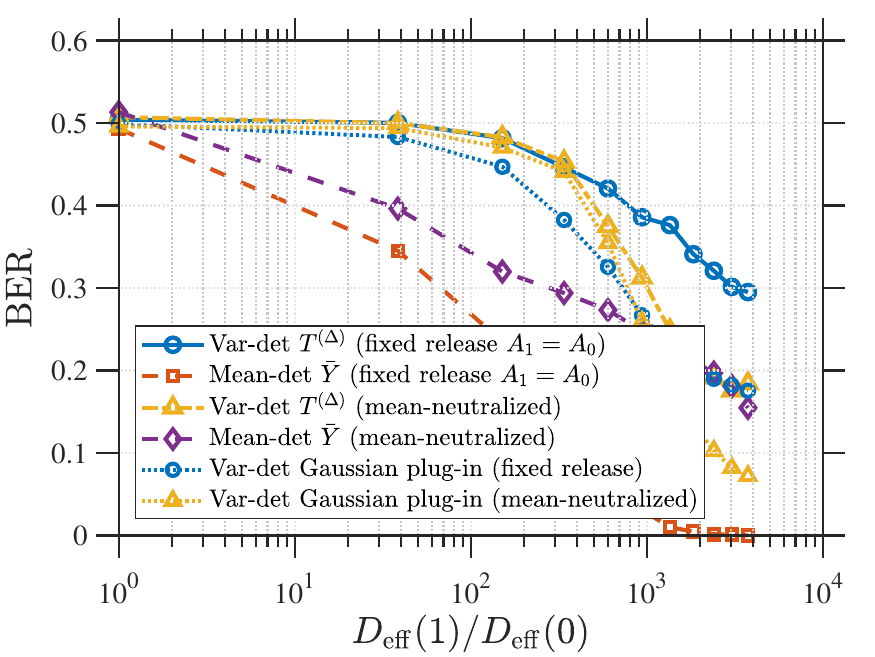}
  \caption{BER versus mobility contrast $D_{\mathrm{eff}}(1)/D_{\mathrm{eff}}(0)$, comparing the proposed dispersion detector $T_k^{(\Delta)}$ with the mean detector $\bar{Y}$ under fixed-release and mean-neutralized settings. The sweep varies only $v(1)$. The quantities $v(0)$, $D_r(0)$, $D_r(1)$, and $D_t$ are fixed. The activity gate is disabled and Gaussian plug-in predictions are fitted from simulated $T_k^{(\Delta)}$ samples.}
  \label{MobilityContrast}
\end{figure}
Fig.~\ref{MobilityContrast} shows how reliability scales with mobility contrast $D_{\mathrm{eff}}(1)/D_{\mathrm{eff}}(0)$ and separates mean-driven gains from dispersion-driven gains. The x-axis is computed from \eqref{eq:Deff_3d_IIC}. Only the symbol-1 speed $v(1)$ is swept, while $v(0)$, $D_r(0)$, $D_r(1)$, and $D_t$ remain fixed. In this sweep, fixed release means $A_1=A_0$, while mean-neutralized means that $A_1$ is recalibrated at each sweep point, using balanced pilot symbols and observable retained-window count means, to approximately match $\mathbb{E}[\bar{Y}_{k,\mathcal{J}}\mid\mathcal{H}_1]$ to $\mathbb{E}[\bar{Y}_{k,\mathcal{J}}\mid\mathcal{H}_0]$ before the main BER run. The activity gate is disabled in this figure, so the BER curves are ungated split-threshold estimates over packets with finite profiled statistics. Near $D_{\mathrm{eff}}(1)/D_{\mathrm{eff}}(0)\!\approx\!1$, all methods operate close to the non-informative regime (BER near $0.5$), indicating that negligible mobility contrast yields little statistical separability. As the contrast increases,
BER decreases for both $T_k^{(\Delta)}$ and $\bar{Y}$, consistent with increasingly distinct counting laws.

The contrast between the two signaling settings highlights the role of overdispersion. Under fixed release $A_1=A_0$,
$\bar{Y}$ improves rapidly and can reach very low BER, reflecting a growing mean-domain separation. Under the
mean-neutralized setting, $\bar{Y}$ degrades markedly, whereas $T_k^{(\Delta)}$ retains an advantage over the tested
contrast values and improves with mobility contrast, isolating second-order information as the dominant discriminator when mean shifts are
suppressed. The Gaussian plug-in curves use empirical moments fitted from simulated $T_k^{(\Delta)}$ samples and track the BER trend under the same no-gate sampling configuration.

\subsection{Correlation \& ISI}

\begin{figure*}[ht] 
     \centering
     \begin{subfigure}[b]{0.45\textwidth}
         \centering
         \includegraphics[width=\textwidth]{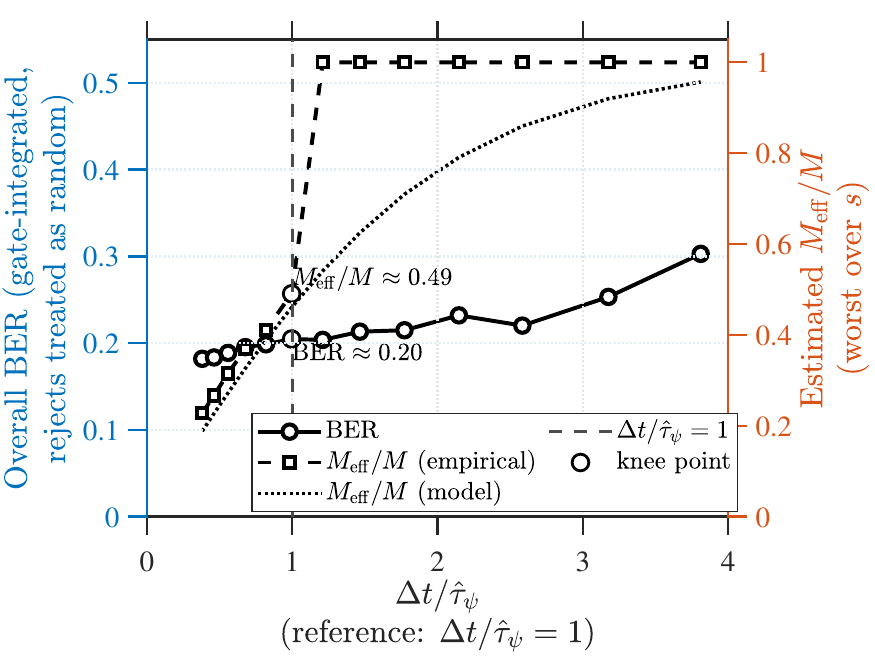}
         \caption{}
         \label{f5}
     \end{subfigure}
     \hfill 
     \begin{subfigure}[b]{0.45\textwidth}
         \centering
         \includegraphics[width=\textwidth]{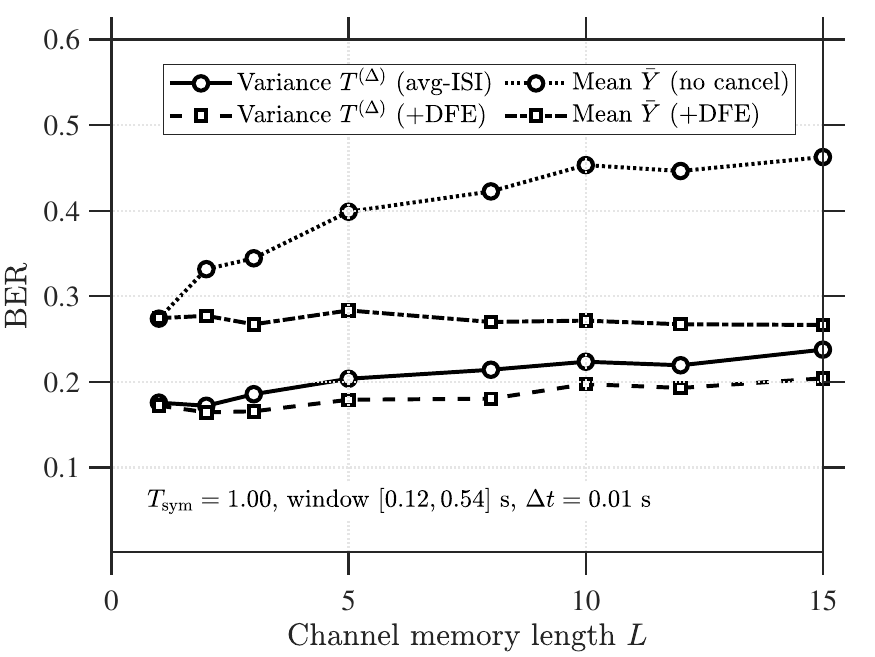}
         \caption{}
         \label{f6}
     \end{subfigure}
     \caption{Dependence and ISI effects. (a) Estimated correlation-adjusted sample ratio $M_{\mathrm{corr}}/M$ and BER versus normalized sampling step $\Delta t/\widehat{\tau}_\psi$. (b) BER versus channel memory length $L$, with and without DFE, for variance and mean detectors.}
     \label{CorrelationISI}
\end{figure*}

Fig.~\ref{CorrelationISI} examines two practical impairments that are not captured by an i.i.d.\ sampling idealization:
within-window correlation of the profiled-residual sequence and residual ISI under a finite channel memory length $L$.
Here $\Delta t$ is the within-window sampling step and $\widehat{\tau}_\psi$ is the estimated correlation time of the
profiled contribution sequence. $M_{\mathrm{corr}}$ denotes the HAC-based, correlation-adjusted effective
independent-sample count for this diagnostic and is distinct from the retained-window length
$M_{\mathrm{eff}}=|\mathcal{J}|$ used in the statistic definition. The BER in Fig.~\ref{CorrelationISI}(a) is
gate-integrated, with gate rejections treated as random decisions.
In Fig.~\ref{CorrelationISI}(a), decreasing the normalized step $\Delta t/\widehat{\tau}_\psi$ increases dependence and
reduces the effective sample ratio $M_{\mathrm{corr}}/M$, with the empirical estimate closely following the proposed
model curve in the tested regime. The marked ``knee'' around $\Delta t/\widehat{\tau}_\psi\approx 1$ corresponds to
$M_{\mathrm{corr}}/M\approx 0.49$ and a representative operating BER near $0.20$, providing a simple rule-of-thumb: sample
spacing much smaller than the correlation time yields diminishing returns because additional samples contribute little
new information, whereas pushing $\Delta t/\widehat{\tau}_\psi$ well above one reduces the number of samples available
in the fixed window and can increase BER.

Fig.~\ref{CorrelationISI}(b) stresses residual ISI by increasing the assumed memory length $L$ and compares no-cancel and
DFE-enhanced variants. The mean detector without cancellation degrades over the tested $L$ range, indicating substantial
bias from accumulating interference. Adding DFE mitigates this trend and stabilizes the mean detector, but it remains
worse than the dispersion-domain curves over the tested range. For $T_k^{(\Delta)}$, performance degrades
more gradually with $L$, and DFE provides a modest additional gain, suggesting that dispersion-based detection is less
sensitive to long-memory interference in this simulation setting while still benefiting from decision-directed cancellation when ISI is severe.

\section{Conclusion}
This paper studied dispersion-domain binary detection in diffusion-based mobile molecular communication under multiplicative geometry uncertainty and time-varying transmitter--receiver separation. Using a conditional Poisson counting model with latent random intensity, we showed how mobility and geometry uncertainty can induce extra-Poisson variability in within-symbol counts. By subtracting the intrinsic Poisson contribution and normalizing by the squared profiled mean, $T_k^{(\Delta)}$ targets this excess dispersion and supports threshold stability under the stated gain model.

The population scaling result is narrow: it applies to pure multiplicative scaling of the signal component. In the implemented receiver, finite samples, profiled mean estimation, the activity gate, additive background, residual dependence, and finite-memory ISI leave residual gain dependence. In particular, the Poisson profile fit generally contributes a first-order score term unless a separate small-sensitivity condition holds. The operational claim is therefore empirical threshold stability over the tested multiplicative-gain regime, rather than exact gain invariance of the implemented receiver. We developed a Gaussian working characterization motivated by a time-series CLT, including long-run-variance dependence correction, gate-integrated false-alarm calibration, ROC/BER approximations, and Gaussian separability design metrics. The simulations support the overdispersion mechanism and show empirical false-alarm stability and reliability gains in the tested settings.

The main limitations are the low-count regime, additive background not well absorbed by profiling, strong mismatch to the multiplicative-gain assumption, finite-sample effects, and residual dependence/ISI. Future work includes joint mean--dispersion fusion, improved composite detectors, online adaptation of the sampling window and dependence parameters, experimental validation, and sequence detection under stronger nuisance dynamics.

\appendices

\section{Oracle Gain Stability and Profiled Statistic}\label{app:oracle_gain_stability}

This appendix records the oracle material underlying the gain-stability discussion in
Sec.~\ref{sec:perf_oracle_gain_stability}: the exact oracle moment identities and the first-order effect of
the implemented Poisson profile fit.

Let $n\triangleq M_{\mathrm{eff}}$ and keep $p$ fixed. Under $\mathcal H_0$, suppose the retained-window latent intensity
admits the factorization
\begin{equation}\label{eq:oracle_scaling_main}
\begin{aligned}
\Lambda_{n,m}^{(\Psi)} &= c_n \Psi\,\widetilde{\Lambda}_{n,m},\\
\Psi &\in\mathcal I=[\Psi_{\min},\Psi_{\max}],\\
0&<\Psi_{\min}<\Psi_{\max}<\infty.
\end{aligned}
\end{equation}
Define
\begin{equation}
\mu_{n,m}^{(\Psi)}\triangleq \mathbb E[Y_{n,m}\mid \mathcal H_0,\Psi]
= c_n\Psi\,\widetilde{\mu}_{n,m},
\qquad
\widetilde{\mu}_{n,m}\triangleq \mathbb E[\widetilde{\Lambda}_{n,m}],
\end{equation}
Let $\widetilde{\kappa}_{r,n,m}$ denote the $r$th cumulant of $\widetilde{\Lambda}_{n,m}$ for $r=2,3,4$.
Define also
\begin{equation}\label{eq:psi_oracle_main}
\psi_{n,m}^{\circ}
\triangleq
\frac{(Y_{n,m}-\mu_{n,m}^{(\Psi)})^2-Y_{n,m}}
{(\mu_{n,m}^{(\Psi)})^2},
\qquad
T_n^\circ \triangleq \frac{1}{n}\sum_{m=1}^{n}\psi_{n,m}^{\circ}.
\end{equation}

\subsection{Exact oracle identities}

\begin{proposition}[Exact oracle identities]\label{prop:oracle_exact_identities}
Under the conditional Poisson model and the notation of Sec.~\ref{sec:perf_oracle_gain_stability}, the oracle contribution in \eqref{eq:psi_oracle_main} satisfies
\begin{align}
\mathbb E[\psi_{n,m}^{\circ}]
&=
\frac{\widetilde{\kappa}_{2,n,m}}{\widetilde{\mu}_{n,m}^{2}},
\label{eq:oracle_mean_main}
\\
\Cov(\psi_{n,i}^{\circ},\psi_{n,j}^{\circ})
&=
\frac{1}{\widetilde{\mu}_{n,i}^2\widetilde{\mu}_{n,j}^2}
\Cov\!\Big(
(\widetilde{\Lambda}_{n,i}-\widetilde{\mu}_{n,i})^2,\nonumber\\
&\qquad\qquad
(\widetilde{\Lambda}_{n,j}-\widetilde{\mu}_{n,j})^2
\Big),
\qquad i\neq j,
\label{eq:oracle_offdiag_main}
\end{align}
so the oracle mean and all nonzero-lag covariances are exactly independent of $\Psi$. Moreover, for the diagonal term,
\begin{equation}\label{eq:oracle_diag_main}
\begin{aligned}
\Var(\psi_{n,m}^{\circ})
&= \frac{1}{\widetilde{\mu}_{n,m}^{4}}
\Bigg[
2\widetilde{\kappa}_{2,n,m}^{2}+\widetilde{\kappa}_{4,n,m}\\
&\qquad
+\frac{4\big(\widetilde{\mu}_{n,m}\widetilde{\kappa}_{2,n,m}+\widetilde{\kappa}_{3,n,m}\big)}{c_n\Psi}\\
&\qquad
+\frac{2\big(\widetilde{\mu}_{n,m}^{2}+\widetilde{\kappa}_{2,n,m}\big)}{(c_n\Psi)^2}
\Bigg].
\end{aligned}
\end{equation}
Hence the lag-zero variance is asymptotically gain-independent as $c_n\to\infty$.
\end{proposition}

\paragraph*{Proof.}

Fix a retained-window index $m$ and suppress the $(n,m,\Psi)$ subscripts for readability. Let
\[
\begin{alignedat}{2}
\mu &\triangleq \mathbb E[Y]=\mathbb E[\Lambda], &\qquad
v &\triangleq \kappa_2(\Lambda),\\
k_3 &\triangleq \kappa_3(\Lambda), &\qquad
k_4 &\triangleq \kappa_4(\Lambda),
\end{alignedat}
\]
and define
\[
\eta^\circ \triangleq (Y-\mu)^2-Y,\qquad \psi^\circ\triangleq \eta^\circ/\mu^2.
\]

\paragraph*{Mean.}
By the mixed-Poisson variance identity,
\[
\mathbb E[\eta^\circ]=\Var(Y)-\mathbb E[Y]=(\mu+v)-\mu=v,
\qquad
\mathbb E[\psi^\circ]=v/\mu^2 .
\]
With $\Lambda=c_n\Psi\,\widetilde{\Lambda}$, $\mu=(c_n\Psi)\widetilde{\mu}$ and
$v=(c_n\Psi)^2\widetilde{\kappa}_2$, giving \eqref{eq:oracle_mean_main}.

\paragraph*{Off-diagonal covariance.}
For $i\neq j$, conditional independence of the counts given the latent intensity vector yields
\[
\Cov(\eta_i^\circ,\eta_j^\circ)
=
\Cov\!\Big(
\mathbb E[\eta_i^\circ\mid \Lambda_i],\,
\mathbb E[\eta_j^\circ\mid \Lambda_j]
\Big).
\]
Since $\mathbb E[\eta^\circ\mid\Lambda]=(\Lambda-\mu)^2$, division by $\mu_i^2\mu_j^2$ gives
\[
\Cov(\psi_i^\circ,\psi_j^\circ)
=
\frac{
\Cov\!\Big(
(\Lambda_i-\mu_i)^2,\,
(\Lambda_j-\mu_j)^2
\Big)
}{
\mu_i^2\mu_j^2
}.
\]
Substituting $\Lambda_\ell=(c_n\Psi)\widetilde{\Lambda}_\ell$ and $\mu_\ell=(c_n\Psi)\widetilde{\mu}_\ell$ for $\ell=i,j$
shows that the numerator and denominator both scale as $(c_n\Psi)^4$, which proves \eqref{eq:oracle_offdiag_main}.

\paragraph*{Diagonal variance.}
Let $X\triangleq Y-\mu$. Then
\[
\eta^\circ = X^2-X-\mu,
\]
and hence
\[
\Var(\eta^\circ)
=
\mathbb E[X^4]-\mathbb E[X^2]^2+\mathbb E[X^2]-2\mathbb E[X^3].
\]
For a mixed-Poisson count,
\[
\begin{aligned}
\kappa_2(Y)&=\mu+v,\\
\kappa_3(Y)&=\mu+3v+k_3,\\
\kappa_4(Y)&=\mu+7v+6k_3+k_4.
\end{aligned}
\]
Combining these with $\mathbb E[X^4]=\kappa_4(Y)+3\kappa_2(Y)^2$ gives
\[
\Var(\eta^\circ)=2\mu^2+4\mu v+2v^2+2v+4k_3+k_4,
\]
so $\Var(\psi^\circ)=\Var(\eta^\circ)/\mu^4$. Substituting
$\mu=(c_n\Psi)\widetilde{\mu}$, $v=(c_n\Psi)^2\widetilde{\kappa}_2$,
$k_3=(c_n\Psi)^3\widetilde{\kappa}_3$, and
$k_4=(c_n\Psi)^4\widetilde{\kappa}_4$ gives \eqref{eq:oracle_diag_main}. \hfill $\square$

\subsection{Implemented-statistic profiling expansion}

We now record what can be proved for the exact profile family used by the detector. On the retained window, write
\[
x_{n,m}\triangleq [u_{n,m},\,1]^\top,\qquad
\mu_{n,m}(\theta)\triangleq x_{n,m}^{\top}\theta,\qquad
\theta=(a,b)^\top .
\]
The implementation enforces positivity through a log-parameterization, but the local expansion below is the same in
the $(a,b)$ coordinates when the pseudo-true parameter is interior. For each $\Psi\in\mathcal I$, assume there is an
interior $\theta_0^{(\Psi)}$ such that
$\mu_{n,m}(\theta_0^{(\Psi)})=\mu_{n,m}^{(\Psi)}$ for $m=1,\ldots,n$, and let $\widehat\theta_n$ be the Poisson
profile MLE used in \eqref{eq:ab_mle}. Define
\begin{equation}\label{eq:phi_profile_def}
\phi(y,\mu)\triangleq \frac{(y-\mu)^2-y}{\mu^2}
=\frac{y^2}{\mu^2}-\frac{2y+1}{\mu}+1,
\end{equation}
so that, up to the fixed degrees-of-freedom factor,
$T_n^{(\Delta)}=(n-p)^{-1}\sum_{m=1}^n\phi(Y_{n,m},\mu_{n,m}(\widehat\theta_n))$.
The derivatives used in the expansion are
\begin{equation}\label{eq:phi_profile_derivatives}
\frac{\partial \phi}{\partial \mu}
=\frac{2y+1}{\mu^2}-\frac{2y^2}{\mu^3},
\qquad
\frac{\partial^2 \phi}{\partial \mu^2}
=\frac{6y^2}{\mu^4}-\frac{2(2y+1)}{\mu^3}.
\end{equation}
The Poisson score contribution for the fitted mean profile is
\begin{equation}\label{eq:profile_score_def}
\begin{aligned}
s_{n,m}(\theta)
&\triangleq
\frac{\partial}{\partial\theta}
\{Y_{n,m}\log\mu_{n,m}(\theta)-\mu_{n,m}(\theta)\} \\
&=
\left(\frac{Y_{n,m}}{\mu_{n,m}(\theta)}-1\right)x_{n,m}.
\end{aligned}
\end{equation}
Let
\begin{equation}\label{eq:profile_A_g_def}
\begin{aligned}
A_n(\Psi)
&\triangleq
-\frac{1}{n}\sum_{m=1}^{n}
\mathbb E_{\Psi}\!\left[
\frac{\partial s_{n,m}(\theta)}{\partial\theta^\top}
\bigg|_{\theta=\theta_0^{(\Psi)}}\right] \\
&=
\frac{1}{n}\sum_{m=1}^{n}
\frac{x_{n,m}x_{n,m}^{\top}}{\mu_{n,m}^{(\Psi)}} .
\end{aligned}
\end{equation}
and define
\begin{equation}\label{eq:profile_g_def}
\begin{aligned}
g_n(\Psi)
&\triangleq
\frac{1}{n}\sum_{m=1}^{n}
\mathbb E_{\Psi}\!\left[
\frac{\partial}{\partial\theta}
\phi(Y_{n,m},\mu_{n,m}(\theta))
\bigg|_{\theta=\theta_0^{(\Psi)}}\right].
\end{aligned}
\end{equation}
If $v_{n,m}^{(\Psi)}\triangleq \Var(\Lambda_{n,m}^{(\Psi)}\mid\mathcal H_0,\Psi)$, then
\begin{equation}\label{eq:profile_g_explicit}
g_n(\Psi)
=
-\frac{1}{n}\sum_{m=1}^{n}
\frac{\mu_{n,m}^{(\Psi)}+2v_{n,m}^{(\Psi)}}
{\big(\mu_{n,m}^{(\Psi)}\big)^3}\,x_{n,m}.
\end{equation}
This follows by substituting
$\mathbb E[Y_{n,m}]=\mu_{n,m}^{(\Psi)}$ and
$\mathbb E[Y_{n,m}^2]=(\mu_{n,m}^{(\Psi)})^2+\mu_{n,m}^{(\Psi)}+v_{n,m}^{(\Psi)}$
in \eqref{eq:phi_profile_derivatives}.

\medskip
\begin{theorem}[Implemented-statistic profiling expansion]\label{thm:implemented_profile_expansion}
Assume, uniformly over $\Psi\in\mathcal I$, that $A_n(\Psi)$ is nonsingular, the profile MLE is an interior solution
and satisfies the usual Poisson $M$-estimator expansion,
\begin{equation}\label{eq:theta_profile_expansion}
\begin{aligned}
\sqrt n\big(\widehat\theta_n-\theta_0^{(\Psi)}\big)
&=
A_n(\Psi)^{-1}\frac{1}{\sqrt n}\sum_{m=1}^{n}s_{n,m}(\theta_0^{(\Psi)})
+o_p(1),
\end{aligned}
\end{equation}
the empirical gradient satisfies
\begin{equation}\label{eq:empirical_gradient_profile}
\begin{aligned}
&\frac{1}{n}\sum_{m=1}^{n}
\frac{\partial}{\partial\theta}
\phi(Y_{n,m},\mu_{n,m}(\theta))
\bigg|_{\theta=\theta_0^{(\Psi)}} \\
&\qquad{}-g_n(\Psi)=O_p(n^{-1/2}),
\end{aligned}
\end{equation}
and the Taylor remainder for \eqref{eq:phi_profile_def} is $o_p(n^{-1/2})$. Then
\begin{equation}\label{eq:implemented_profile_expansion}
\begin{aligned}
\sqrt n\Big(T_n^{(\Delta)}-T_n^\circ\Big)
&=
g_n(\Psi)^\top A_n(\Psi)^{-1}
\frac{1}{\sqrt n}\sum_{m=1}^{n}s_{n,m}(\theta_0^{(\Psi)}) \\
&\quad
+o_p(1),
\end{aligned}
\end{equation}
uniformly in $\Psi\in\mathcal I$.
\end{theorem}

\paragraph*{Proof.}
Since $p$ is fixed and $n^{-1}\sum_m\phi(Y_{n,m},\mu_{n,m}^{(\Psi)})=O_p(1)$, the difference between the factors
$1/(n-p)$ and $1/n$ is $o_p(n^{-1/2})$. A Taylor expansion of
$n^{-1}\sum_m\phi(Y_{n,m},\mu_{n,m}(\widehat\theta_n))$ around $\theta_0^{(\Psi)}$ gives, with
$D_n(\Psi)$ denoting the empirical gradient in \eqref{eq:empirical_gradient_profile},
\[
\begin{aligned}
T_n^{(\Delta)}-T_n^\circ
&=
D_n(\Psi)^\top
\big(\widehat\theta_n-\theta_0^{(\Psi)}\big) \\
&\quad
+o_p(n^{-1/2}),
\end{aligned}
\]
uniformly in $\Psi$. By \eqref{eq:empirical_gradient_profile} and
\eqref{eq:theta_profile_expansion}, the bracketed empirical gradient can be replaced by $g_n(\Psi)$ at
$o_p(n^{-1/2})$ cost. Substituting the profile-MLE expansion
\eqref{eq:theta_profile_expansion} gives \eqref{eq:implemented_profile_expansion}. \hfill $\square$

\medskip
\noindent\textbf{Sufficient condition for oracle transfer.}
If, in addition to the assumptions of Theorem~\ref{thm:implemented_profile_expansion},
\begin{equation}\label{eq:small_profile_sensitivity}
\sup_{\Psi\in\mathcal I}
\left\|g_n(\Psi)^\top A_n(\Psi)^{-1}\right\|\to 0
\end{equation}
and $n^{-1/2}\sum_m s_{n,m}(\theta_0^{(\Psi)})=O_p(1)$ uniformly in $\Psi$, then
\begin{equation}\label{eq:profiling_remainder_main}
\sqrt n\Big(T_n^{(\Delta)}-T_n^\circ\Big)\xrightarrow{p}0
\qquad\text{uniformly in }\Psi\in\mathcal I.
\end{equation}
In that special case, the oracle CLT for $T_n^\circ$ transfers to the implemented statistic by Slutsky's theorem.

Condition \eqref{eq:small_profile_sensitivity} is an additional small-sensitivity requirement, not a consequence of
the scale-and-offset profile family alone. To see this, consider the one-parameter scale special case
$\mu_{n,m}^{(\Psi)}=c_n\Psi u_{n,m}$ and
$v_{n,m}^{(\Psi)}=(c_n\Psi)^2\widetilde v_{n,m}$ with $u_{n,m}>0$. Use the shorthand
$\overline u_n\triangleq n^{-1}\sum_m u_{n,m}$,
$\overline{u^{-1}}_n\triangleq n^{-1}\sum_m u_{n,m}^{-1}$, and
$\overline{\widetilde v u^{-2}}_n\triangleq n^{-1}\sum_m \widetilde v_{n,m}u_{n,m}^{-2}$. Then
\[
\begin{aligned}
g_n(\Psi)A_n(\Psi)^{-1}
&=
-\frac{(c_n\Psi)^{-1}\overline{u^{-1}}_n}{\overline u_n}
-\frac{2\overline{\widetilde v u^{-2}}_n}{\overline u_n}.
\end{aligned}
\]
Thus even though $g_n(\Psi)=O((c_n\Psi_{\min})^{-1})$ in this high-count scaling, the premultiplied sensitivity
$g_n(\Psi)A_n(\Psi)^{-1}$ is generally $O(1)$ whenever the normalized overdispersion terms
$\widetilde v_{n,m}$ do not vanish. Hence the profiling contribution in \eqref{eq:implemented_profile_expansion}
is generally first-order for the implemented statistic. The zero-remainder claim should not be used without verifying
\eqref{eq:small_profile_sensitivity}.

\section{Window Choice and Residual Dependence}\label{app:window_and_dependence}

This appendix records (i) the calibration rule used to select the within-symbol index set $\mathcal{J}$ and hence
$M_{\mathrm{eff}}$, and (ii) the long-run variance quantities used to account for within-symbol dependence of the
$\{\widehat{\psi}_{k,m}\}$ sequence.

\subsection{Calibration rule for selecting $\mathcal{J}$}

Using the normalized mean-shape template $u(m)$ learned in Sec.~III-B, define the normalized cumulative template energy
\begin{equation}\label{eq:app_Cm_m1m2}
\begin{aligned}
C(m) &\triangleq
\frac{\sum_{j=1}^{m} u(j)}{\sum_{j=1}^{M} u(j)}, \qquad m=1,\ldots,M,\\
m_1 &\triangleq \min\{m:\ C(m)\ge \alpha\},\\
m_2 &\triangleq \max\{m:\ C(m)\le 1-\beta\},
\end{aligned}
\end{equation}
for fixed design constants $\alpha,\beta\in(0,1)$. The retained index set and effective window length are
\begin{equation}\label{eq:app_Mset_Meff}
\mathcal{J}\triangleq \{m_1,\ldots,m_2\},\qquad
M_{\mathrm{eff}}\triangleq |\mathcal{J}|=m_2-m_1+1.
\end{equation}
The only requirement for the statistic in \eqref{eq:Tdelta_def_final} is $M_{\mathrm{eff}}>p$.

\subsection{Long-run variance for dependent $\widehat{\psi}_{k,m}$}

For each symbol $k$, define the within-window average and centered sequence
\begin{equation}\label{eq:app_psicenter}
\begin{aligned}
\bar{\psi}_k &\triangleq
\frac{1}{M_{\mathrm{eff}}}\sum_{m\in\mathcal{J}} \widehat{\psi}_{k,m},\\
\psi^{\mathrm{c}}_{k,m} &\triangleq \widehat{\psi}_{k,m}-\bar{\psi}_k,\qquad m\in\mathcal{J}.
\end{aligned}
\end{equation}
The superscript ``c'' denotes centering and is unrelated to the oracle contribution $\psi^\circ$ in
Appendix~\ref{app:oracle_gain_stability}. Let $\mathcal{K}_s$ denote the set of labeled calibration symbols under
hypothesis $\mathcal{H}_s$. These labeled sets are used for analytical variance/ROC characterization. The detector threshold calibration itself
uses $\mathcal{H}_0$ symbols only as described in Sec.~\ref{sec:det_threshold_stability}.

For lag $\ell\ge 0$, define the pooled normalization
\begin{equation}\label{eq:app_Nsl}
N_{s,\ell}\triangleq |\mathcal{K}_s|\,(M_{\mathrm{eff}}-\ell),
\end{equation}
and estimate the lag-$\ell$ autocovariance by
\begin{equation}\label{eq:app_gammahat_compact}
\widehat{\gamma}_s(\ell)\triangleq
\frac{1}{N_{s,\ell}}
\sum_{k\in\mathcal{K}_s}\,
\sum_{i=1}^{M_{\mathrm{eff}}-\ell}
\psi^{\mathrm{c}}_{k,m_i}\,\psi^{\mathrm{c}}_{k,m_{i+\ell}},
\end{equation}
where $\{m_i\}_{i=1}^{M_{\mathrm{eff}}}$ is the ordered list of indices in $\mathcal{J}$.

A Bartlett (Newey--West) long-run variance estimate is
\begin{equation}\label{eq:app_lrv_compact}
\widehat{\omega}^2_{\psi,s}\triangleq
\widehat{\gamma}_s(0)+
2\sum_{\ell=1}^{L_s}\Big(1-\frac{\ell}{L_s+1}\Big)\widehat{\gamma}_s(\ell),
\end{equation}
and the corresponding dependence inflation factor is
\begin{equation}\label{eq:app_Omega_compact}
\widehat{\Omega}_s \triangleq
\frac{\widehat{\omega}^2_{\psi,s}}{\widehat{\gamma}_s(0)} \ \ge\ 1.
\end{equation}
With the degrees-of-freedom correction in \eqref{eq:Tdelta_def_final} and the usual long-run variance scaling,
\begin{equation}\label{eq:app_varT_compact}
\Var\!\big(T_k^{(\Delta)}\mid \mathcal{H}_s\big)\approx
\frac{M_{\mathrm{eff}}}{(M_{\mathrm{eff}}-p)^2}\,\widehat{\omega}^2_{\psi,s}.
\end{equation}
For moderate/large $M_{\mathrm{eff}}$ (typically $M_{\mathrm{eff}}\!\gg\! p$), we use the convenient simplification
$\Var(T_k^{(\Delta)}\mid \mathcal{H}_s)\approx \widehat{\omega}^2_{\psi,s}/(M_{\mathrm{eff}}-p)$ by taking
$M_{\mathrm{eff}}\approx M_{\mathrm{eff}}-p$.

The truncation lag $L_s$ is selected from calibration and capped by the user-chosen maximum HAC lag $L_{\max}$ by inspecting the sample
autocorrelation $\widehat{\rho}_s(\ell)\triangleq \widehat{\gamma}_s(\ell)/\widehat{\gamma}_s(0)$ and choosing the
smallest lag after which $|\widehat{\rho}_s(\ell)|$ remains below a small threshold for several consecutive lags.

\bibliographystyle{IEEEtran}
\bibliography{references}

\vfill
\end{document}